# Interplanetary Medium as Fuel for Plasma Thrusters

A. R. Karimov[1,2], A. M. Bulygin[1*], P. A. Murad[3], A. E. Shikanov[1], A. P. Skripnik[1]

*1 Department of Electrophysical Facilities, National Research Nuclear University MEPhI, Kashirskoye shosse 31, Moscow, 115409, Russia*
*2 Institute for High Temperatures, Russian Academy of Sciences, Izhorskaya 13/19, Moscow, 127412, Russia*
*3 Morningstar Applied Physics, LLC, Vienna, USA*

*We examine the use of the interplanetary substance as a possible fuel for spacecraft's plasma thrusters for interplanetary flights. The solar radiation is considered as an energy source for the ionization and the acceleration of particles of captured space environment that is used as drop weight. The method of capturing the space environment, which is dependent on its density, the velocity of the spacecraft, and the power of solar radiation, defines the technical feasibility of this scheme. Both pulsed and continuous operations of the plasma accelerator are possible under some conditions as shown in the estimates.*

## Introduction

The exploration of the Solar system requires a creation of spacecraft capable to be accelerated to high speeds. This immediately imposes certain physical and technical restrictions on the method for creating thrust, which follows directly from the Tsiolkovski's formula (see, for example, [1]):

$$v_f = v_{ex}\ln\left(\frac{m_0 + m_f}{m_0}\right), \quad (1)$$

where $v_f$ is the final velocity of the rocket (based upon 'specific impulse'), $m_0$ is the empty mass of the spacecraft, $m_f$ is the mass of the propellant, and $v_{ex}$ is the effective exhaust velocity. The logarithmic equation (1) shows that the increase in the speed of the rocket is possible only by accelerating the jet (for which there are many methods), while not increasing the mass of its propellant.

The maximum speed of chemical engines is only about $10^3$ m/s, the maximum speed of nuclear engines is approximately $10^4$ m/s, while the end-Hall thrusters are capable of exhaust velocity in the order of $5\times 10^5$ m/s [1—3], i.e. plasma thrusters have the advantage over other types of engines at least in terms of of the feasible specific impulse. This makes this type of engine the most attractive for long-distance space flights. Also, the potential capabilities of chemical and nuclear engines are in principle limited due to the energy released in the corresponding chemical and nuclear reactions. By contrast, plasma thrusters do not have these limitations due to an external energy supply. Thus, the energy of the plasma flow is limited only by the acceleration method and the value of the input energy.

Fundamentally new approaches to possible technical solutions appear. For example, there are several schemes of plasma thrusters among Hall-effect thrusters such as end thrusters, stationary Morozov thruster, and Zharinovsky thruster with an anode layer [2—8]. The differences between these devices are in the approach to generating and accelerating the plasma flow. This is reflected in the thrust magnitude, which is the integral technical characteristic of the engine $F = \dot{m}v_{ex}$, where $\dot{m}$ is the fuel consumption rate. As shown from this ratio, the higher the speed of the flow the less fuel is required to create the desired thrust. Unfortunately, this also reveals the main disadvantage of plasma thrusters based upon a very small technically available value of the mass $\dot{m}$, i.e. modern plasma engines provide a thrust of the order of Newtons (see, for example, [2,9]).

However, the thrust of the plasma thruster can be created by an external energy source. Solar radiation can be such a source, energy from which is spent on ionization of the environment to create thrust. Theoretically, the interplanetary space medium could be used as an operating fuel. This idea has been discussed for decades, starting with Bussard's work on a a space ramjet as well as a thermonuclear spacecraft engine (see, for example, [10—15]). However, it is potentially possible to increase the engine thrust only by increasing the specific impulse rather than considering the small value of $\dot{m}$, which is extracted from outer space. This is feasible only in plasma accelerators.

Thus, it would be of interest to estimate the feasibility of using the interplanetary medium for creating and accelerating plasma flows to distant flights in the Solar system. This implies solving two interrelated problems. The first problem is how to extract the necessary amount of interplanetary medium for plasma ionization and the second is to accumulate solar energy to provide ionization and acceleration of the generated plasma with acceptable values of spacecraft's speed and thrust. In the present paper, these issues will be addressed.

## 2. The acceleration of variable mass objects taking into account capture and ejection of particles

Let us explain the acceleration dynamics of an object with variable mass, taking into account the input flow of additional particles coming from the interplanetary medium and ejected particles after acceleration, to demonstrate the acceleration scheme. We consider conditions that are possible to neglect the influence of the external forces, for example at places, where the effects of deceleration are significant (far away from any centers of gravity or outside the atmosphere). To do this, we use the generalized Meshchersky equation without treating external forces [16]:

$$m_0 \frac{d\boldsymbol{v}}{dt} = \boldsymbol{u}_+ \frac{dm_+}{dt} - \boldsymbol{u}_- \frac{dm_-}{dt}, \qquad (2)$$

where $\boldsymbol{v}$ is the spacecraft speed in the inertial frame of reference, $m_+$ is the mass of particles captured by the spacecraft and $m_-$ is the mass of particles thrown out into the environment. In this case, the dependence shows the link between the speeds $u_+$ and $u_-$ of the captured and ejected mass relative to the spacecraft and its corresponding absolute values:

$$\begin{cases} \boldsymbol{v}_+ = \boldsymbol{v} + \boldsymbol{u}_+, \\ \boldsymbol{v}_- = \boldsymbol{v} + \boldsymbol{u}_-. \end{cases} \qquad (3)$$

Neglecting the initial velocity of particles, we assume that the particles are captured and thrown only along the same direction of the spacecraft's motion. For simplicity, we study the case when the entire interplanetary medium is stationary, i.e. $v_+ = 0$ where this is the velocity of the medium in the laboratory frame of reference. As seen from the first relation of the system (3), $u_+ = -v$, the speed of the dust flow that hits the surface of the spacecraft. Let us assume that the velocity of the ejected particles relative to the spacecraft's speed is constant and equals $u_- = -u_{ex}$. Here, the absolute value of $u_{ex}$ is determined by the acceleration mechanism, which is beyond the study. Nevertheless, in general, $u_{ex}$ could be relativistic, i.e. the mass of ejected particles should be taken in the form:

$$\mu_- = \frac{\mu_0}{\sqrt{1 - \left(\frac{u_{ex}}{c}\right)^2}}, \qquad (4)$$

where $\mu_0$ is the interplanetary particle's rest mass.

The equation of motion (2) in the assumptions made can be written in the scalar form:

$$m_0 \frac{dv}{dt} = -v \frac{dm_+}{dt} + u_{ex} \frac{dm_-}{dt}. \qquad (5)$$

Here the attached mass is created only by an income flow of dust particles.

$$\frac{dm_+}{dt} = \mu_0 \, S \, n_{im} \, u_+ = \mu_0 S n_{im} v,$$

where S is the characteristic square of a cross-section through which particles of the interplanetary medium reach the surface of the spacecraft (the characteristic square of a trap for interplanetary substance, which is shown in Fig. 2, is the area of the trap inlet), $n_{im}$ is the density of the captured interplanetary medium. Assume that ejected mass $dm_-/dt$ is the accelerated flow of income particles we can write

$$\frac{dm_-}{dt} = \frac{\mu_0 \, S \, n_{im} \, v}{\sqrt{1 - \left(\frac{u_{ex}}{c}\right)^2}} = \frac{1}{\sqrt{1 - \left(\frac{u_{ex}}{c}\right)^2}} \frac{dm_+}{dt}.$$

As a result, the equation (5) can be written

$$\frac{dv}{dt} = \alpha(v_{ex} - v)v,$$

where $\alpha = S \, n_{im} \, \mu_0/m_0$ and $v_{ex} = u_{ex}/\sqrt{1 - (u_{ex}/c)^2}$. The solution of this equation with the initial condition $v(t = 0) = v_0$ is

$$v = \frac{v_0 v_{ex} \exp(t/\tau)}{v_{ex} + v_0(\exp(t/\tau) - 1)}, \qquad (6)$$

where

$$\tau^{-1} = \alpha v_{ex} = \frac{\mu_0}{m_0} S \, n_d \frac{u_{ex}}{\sqrt{1 - \left(\frac{u_{ex}}{c}\right)^2}}. \qquad (7)$$

This equation defines the characteristic acceleration time depending on the accelerator and the interplanetary medium. The fundamental difference between this formula and the dependence (1) is that here time is directly included in equation (6). Note that the relation (6) has a limit $v_{ex}$, unlike equation (1), which has no speed limit. As is seen, the required acceleration time depends directly on the value S, methods of plasma acceleration, its components, and density, i.e. it depends on the characteristics of the interplanetary medium and methods of the plasma capture.

## 3. Characteristics of the interplanetary medium

We will analyze the physical and chemical characteristics of the interplanetary medium of the Solar system, which are of interest from the usage of this medium as fuel for plasma thrusters. It is acceptable to neglect the interstellar component, taking into account the electromagnetic radiation of the Sun only, even though the interplanetary medium is seen as the entire set of matter and fields which fill interplanetary space inside the Solar system. As the main part of interplanetary matter, we will consider only the solar wind, high-energy charged particles, interplanetary dust, and neutral gas [17-25].

The solar wind is a flow of hydrogen plasma with a negligible fraction of helium ions (several percent of the total number of particles), which is generated by transfer processes in the solar corona and inside the interior of the Sun itself. Direct observations [17-22] have shown that non-stationary, spherically asymmetric flows are formed, which differ greatly from each other and change over time depending on the state of the solar corona. Plasma outflow occurs with the formation of non-stationary shock waves in interplanetary space near the Sun, characterized by sharp jumps in speed, a density, and temperature when faster flows catch up with slower ones.

As they move away from the Sun, these waves interact with each other, forming a quasi-stationary plasma flow at a distance of about 1 Astronomical Unit (A.U.). A relatively slow part of the solar wind is constantly present in this flow, with a characteristic speed of 300 - 500 km/s and a density of 10 - 15 cm$^{-3}$ [22,24]. This part of the flow appears during the gas-dynamic expansion of the "quiet" part of the solar corona. A quasi-stationary (fast) flow has a speed of 700 - 1000 km/s and a density of 3 - 4 cm$^{-3}$. It is formed by particles that fly directly from the surface of the Sun [22,24]. This phenomenon is observed for several months and repeats every 27 days when viewed from the Earth. This period equals to the period of the rotation of the Sun. Another component is the relativistic flow of protons generated in the solar atmosphere during solar flares, but due to its short duration and low density, its contribution can be ignored [22,24]. As a result, a quasi-stationary plasma's flow with an average speed of $v_{sw}$ = 400 - 500 km/s and a density, that lies in the range of $10 \leq n_{sw} \leq 20$ cm$^{-3}$, is fixed in the Earth's orbit (at a distance of r ~1 A.U.) [22,24,25].

The pressure gradient and the force of gravity of the Sun can be ignored at a distance of about 1 A.U., which follows from the calculations [22]. This means that the wind speed at distances r > 1 A.U. can be considered constant, and the flow density decreases with the distance as r$^{-2}$. For example, at a distance of $r = 2$ A.U., the density is in the range $2 \leq n_{sw} \leq 5$ cm$^{-3}$. We will take the average value of density of $n_{sw} \cong 9$ cm$^{-3}$, which is calculated near $r = 1$ A.U. and $r = 2$ A.U., as a characteristic value of the solar wind density in a range of distances $1 \leq r \leq 2$ A.U. It is obvious that when $r < 1$, this value will increase as we approach the Sun.

Interplanetary dust fills the entire heliosphere very irregularly and is mainly concentrated near the Sun, because of the solar radiation pressure and the Sun's gravitational force affects differently on particles depending on their size and charge. Light particles with a mass of $10^{-15} < m < 10^{-13}$ g will be removed from the solar neighborhood along straight paths because of the influence of light pressure. One part of the large particles that have a spherical shape will settle on the surface of the Sun. Another part moves along spiral trajectories, that gets away from the Sun due to the Poynting-Robertson effect [20,21,26].

As a result of this impact, about 90% of nanometer particles (m $\leq 10^{-18}$ kg) are trapped near the Sun and the remaining part is injected from a distance of 0.15 A.U. filling the space up to $r \leq R_* = 3$ A.U. (dust is practically absent at distances of $R_* > 3$ A.U. [19,23,27]). Observations have shown that the main amount of dust is concentrated in the plane of the Ecliptic, which is sometimes called the zodiacal dust cloud. This is caused by the collective action of the gravitational forces of the solar system planets rotating in this plane. This spatial distribution was observed by Helios 1 and 2 space probes at distances from 0.3 to 1 A.U. from the Sun [20]. It was found that the dust concentration decreases in ecliptic as a function of distance from the Sun as r$^{-1.3}$. This spatial relationship leads to the fact that the dust density at a distance of $r = 2$ A.U. will decrease approximately two times compared to the dust density near the Earth.

The interplanetary dust consists of many chemical elements such as Mg, Si, Fe, S, Co, Os, Ir, Ni, Al, Cr, Mn, which are found in silicates, metal oxides, sulfur and carbon compounds. The first four elements are mostly in the dust [27,28]. There is also a small amount of neutral hydrogen and noble gases that penetrate the dust particles from the interstellar medium. However, there is currently no information on the full quantitative content of elements in dust particles.

Therefore we will neglect the relative content of elements in dust particles and will make a rough estimate of the maximum density of dust particles n$_{max}$ for the lightest dust element Mg ($\mu_{Mg} = 40 \times 10^{-24}$ kg) and the minimal density of dust particles n$_{max}$ for the heaviest dust element Fe ($\mu_{Mg} = 93,5 \times 10^{-24}$ kg), using information about the annual amount of dust coming from the zodiac cloud into the earth's atmosphere as it rotates around the Sun. According to observations [26-27], $M_d = 25 \div 45$ thousand tons of the interplanetary dust is settled on the Earth per year, which corresponds to a mass flow of $J_d = 0,79 \div 1,42$ kg/s. In further estimates, we will take the mass flow of the dust of $J_d = 1.11$ kg/s as a characteristic value of the mass of the dust flow falling on the Ground. This value equals the average of the mass flow. On the other hand, this flux is expressed by a ratio $J_d = \mu_d n_d U S_E$, where $\mu_d$ is the mass of a dust particle, $n_d$ is the dust density, U is the speed of the dust flow relative to the Earth, $S_E$ is an effective cross-section of the Earth through which dust flow enters the Earth's atmosphere. We will take the speed of the Earth's rotation around the Sun $U = v_E \approx 30$ km/s as the speed of the flow, assuming that the dust is stationary and the effective cross-section is a circle with radius $R_{ef} = R_E + h$, where $R_3 \approx 6371$ km is the radius of the Earth and h ≈ 1000 km is the height of the atmosphere, above which there are no collisions with particles. Then the density of dust near the Earth according to the given astronomical data, estimated by the ratio $n = J_d/(\pi \mu_d v R_{ef}^2)$, has the minimum value of $n_{Fe}$ = 2.32 cm$^{-3}$ and the maximum value of $n_{Mg}$ = 5.41 cm$^{-3}$. It should be noted that these results are within an order of magnitude the same as the experimental values [23], for which a density of the dust was $n_p$ = 10 cm$^{-3}$ in near-earth orbit.

From this estimation, it follows that the uncertainty in the composition of interplanetary dust leads to a change in the density value making this a very inhomogeneous region. This change is comparable with the decrease in density, which is observed at a distance of 1 A.U. from the Earth. For example, the dust density changes by only 12% at a distance of $\Delta r = 0.1$ A.U. from the Earth, while the uncertainty of the composition can lead to an uncertainty of the density value of the order of 50%. Considering this fact we will neglect the spatial dependence of the dust density in the present estimations for the region $1 \leq r \leq 2$ A.U. and will equate it to a certain constant from the range $2{,}32 < n_d < 5{,}41$ cm$^{-3}$.

Neutral gas in the Solar system was discovered by observing resonantly scattered solar radiation [20,26]. According to these observations, the density of hydrogen atoms is 0.06 cm$^{-3}$ and the density of helium atoms is 0.012 cm$^{-3}$, it means the density of neutral gas is much smaller than the density of the solar wind and dust, and it can be neglected.

Thus, only solar wind particles and interplanetary dust can be used as a fuel for plasma thrusters. At a distance of about 1 A.U. these components provide the density of the interplanetary medium $n = n_{sw} + n_d \cong 12 \div 15$ cm$^{-3}$ [23], but it is necessary to take into account the magnitude and direction of the solar wind flow relative to the ship if the local density is being determined. We will consider the case when only stationary interplanetary dust is used as fuel with the density of $n_d \cong 2.32 \div 5.41$ cm$^{-3}$ to simplify the estimates and to ignore this process. It should be noted that all obtained results for these interplanetary medium densities give very rough estimates of considered processes. These estimates should be interpreted as conditions that show the feasibility of the discussed scheme.

We will estimate the radiation power distribution as a function of the distance from the Sun to find out the possibility of using solar radiation for ionizing neutral atoms and their subsequent acceleration. Taking into account that the surface temperature of the Sun is $T_C$ = 6000 K at a distance of $R_C = 6.9 \times 10^8$ m [20, 22], we get the radial power distribution from the Stephan-Boltzmann's law:

$$W_C = \frac{\sigma T_C^4}{(r/R_C)^2} \qquad (8)$$

where $\sigma$ is the Stefan-Boltzmann constant. This dependence is sketched in Fig. 1, where for the convenience of perception, the distances from the Sun to the planets falling into a sphere of radius $R*$ ($R*$ is the distance from the Sun to the edge of the asteroid belt, beyond which the dust is almost not present) are labeled.

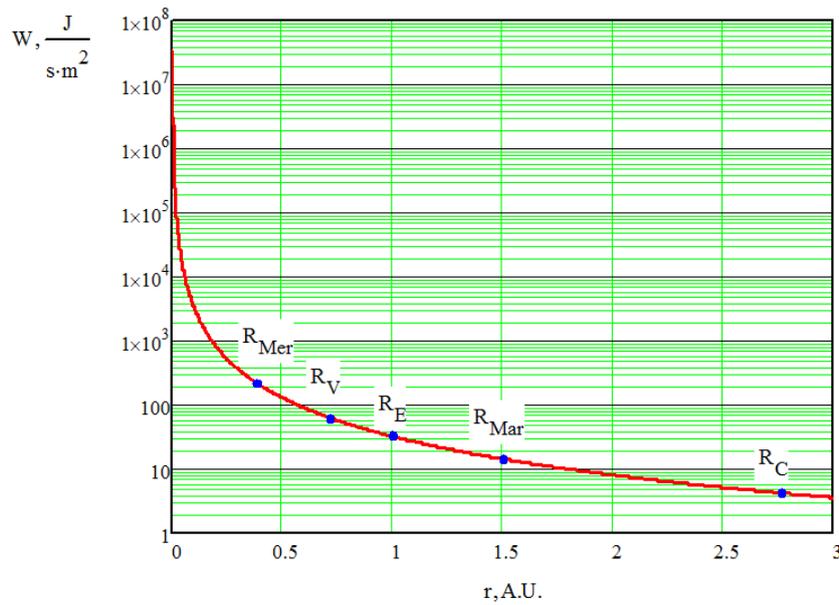

**Fig. 1.** The radiation power with respect to the distance from the Sun (see eq. (8)). Here $R_{Mer}$ corresponds to the Mercury orbit, $R_V$ – the Venus orbit, $R_E$ - the Earth orbit, $R_{Mar}$ - the Mars orbit and $R_C$ - the Ceres orbit.

## 4. Accumulation of a substance from the interplanetary medium

Presented data about the distribution of interplanetary medium in the Solar system clearly shows that it is impossible to directly use the interplanetary medium as fuel for future plasma thruster. Therefore, based on this information, it is necessary to discuss possible ways of increasing the medium density to levels sufficient for producing a low-pressure discharge. Also, it is necessary to evaluate the parameters of plasma, which can be used in a plasma accelerator by using the data of solar radiation distribution.

The simplest way to increase the ionizing medium is by using a conical trap (see, Fig. 2). Such a conical structure should be placed on a moving spacecraft in the direction of its travel. The particles of the incoming flow are repeatedly reflected from the conical surface and fall into the region of the discharge chamber. Proceeding from the continuity of the particle flow one can conclude that the particle's density in the narrow part of the conical trap can be increased concerning the input density n in k =$D^2/d^2$ times, where k is the ratio of the area of the inlet and discharge chamber, D is the inlet's diameter of the conical trap, and d is the input diameter of the discharge chamber. For example, for the sizes, D = 20 km and d = 20 cm of the trap shown in Fig. 2, we obtain k = $10^{10}$ only due to the geometry of the trap. This free-molecular mode is feasible as long as the medium remains collisionless even in the narrow part of the trap, i.e. the free path length calculated from the density in the discharge chamber area needs to be at least comparable to the characteristic size of the device (e.g. the diameter of trap's inlet).

Moreover, sufficiently high speed of the spacecraft is required that this method works, since the accumulation time of particles in the discharge chamber is inversely proportional to the speed of the spacecraft and proportional to its volume, i.e. the technical description is needed.

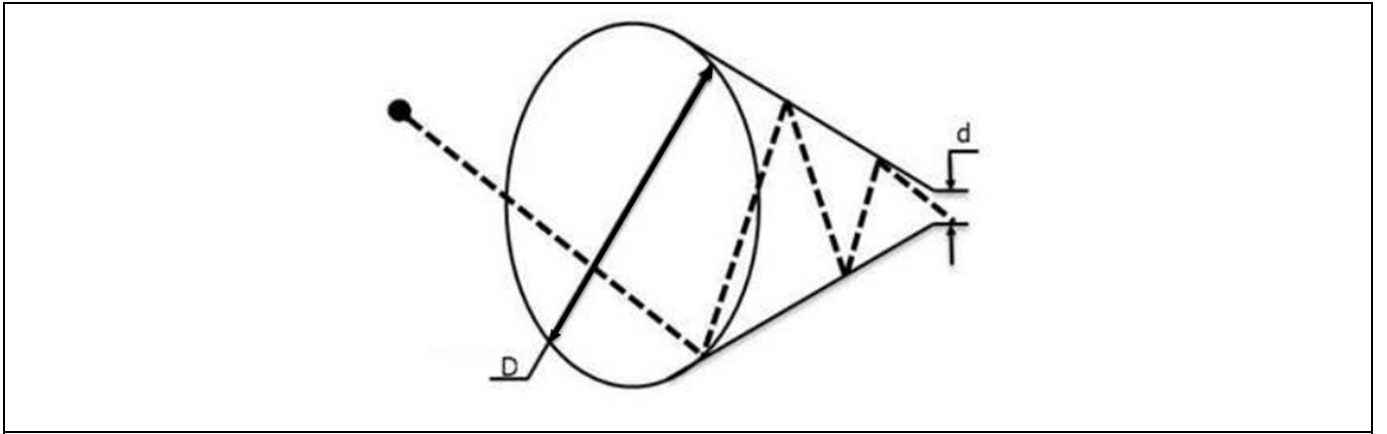
**Fig. 2.** The scheme of the trap for interplanetary substance (here, we set D = 20 km and d = 20 cm).

Besides, sufficiently high speed of the spacecraft is needed to make this method work, since the time of accumulation of particles in the discharge chamber is inversely proportional to the speed of the spacecraft and proportional to the volume of the discharge chamber, i.e. some technical detail is required. As an example, we will consider a cylindrical discharge chamber with a volume of $V = 10^3$ cm$^3$, where a low-pressure discharge is initiated. As a rule, the RF field can lead to such types of discharges at a pressure of $p\sim10^{-1}$ Pa (see, for example, [2,30,31]) that corresponds to the particle density $n_0 \sim 10^{10} - 10^{11}$ cm$^{-3}$ inside the discharge chamber under normal conditions. Thus we put the density in the discharge chamber being equal to the desired value of $n_0 = 10^{11}$ cm$^{-3}$ in estimations of the mean free path $\lambda$. Then we get $\lambda = 1/(\sigma_g n_0) = 10^5$ cm for the characteristic gas-kinetic cross-section $\sigma_g = 10^{-16}$ cm$^2$, that is much larger than the characteristic size of the discharge chamber, so there is no need to take into account the initial pressure of the medium, hence the collisionless regime is provided.

The speed of the particles entering the trap relative to the ship matches the instantaneous speed of the spacecraft (see section 2) since only stationary interplanetary dust was selected as fuel. In this case, the incoming flux of the dust particles will have filled the selected volume V to a given number of particles $N_0$ for the time $\tau_{acc}$:

$$\tau_{acc} = \frac{4N_0}{\pi D^2 v n_d}, \qquad (9)$$

where $N_0 = V n_0$ and $n_d$ are previously calculated values of densities of the captured interplanetary medium of $n_{Fe} = 2.32$ cm$^{-3}$ and $n_{Mg} = 5.41$ cm$^{-3}$ for the cases, then the flux is created only by the "light" or "heavy" dust particles (Mg and Fe), respectively.

In Fig. 3 the characteristic times of particles' accumulation for these values are plotted as a function of the spacecraft speed, which gives an idea of the time scale of the process.

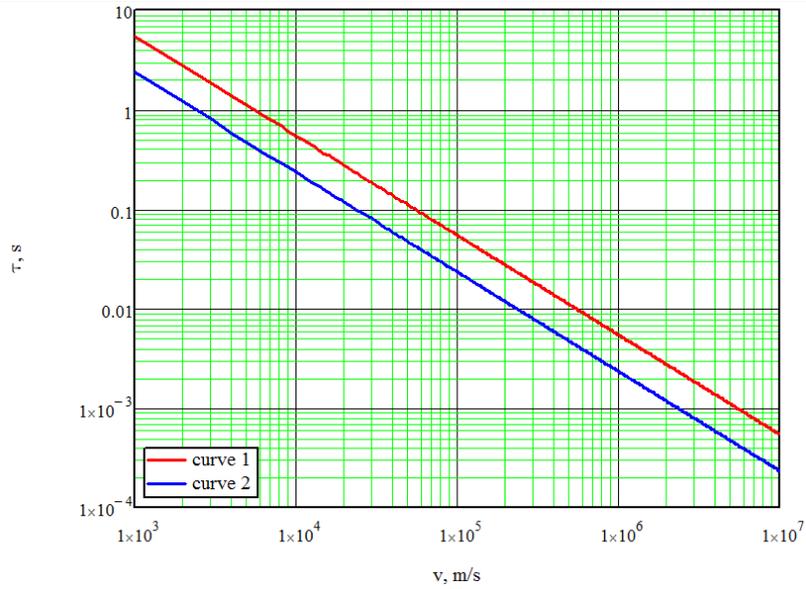

**Fig. 3.** Accumulation time as a function of the relative velocity for the incoming flow of the dust particles in the logarithmic scale (Curve 1, 2 correspond densities of $n_{Fe} = 2.32\times10^{10}$ cm$^{-3}$ and $n_{Mg} = 5.41\times10^{10}$ cm$^{-3}$, respectively).

## 5. The estimation of the thrust generated by the attached mass

We will demonstrate the efficiency of the scheme by pre-evaluating the resulting thrust and will assume that the entire captured particle flow is ionized and accelerated to the required energy. The payload has a mass of $m_0 = 1T = 10^6$ g and the trap has used earlier in this study, the outer diameter D which gives $S = 3.14\times10^8$ m$^2$. We nominally consider that the whole dust consists of only "heavy" atoms of iron or only "light" atoms of magnesium with corresponding densities. Let's assume the initial speed of the spacecraft is equal to $v_0 = 10^3$ m/s. The speed of the accelerated flow is assumed to be the characteristic speed of $u_{ex} \approx v_{ex} = 10^7$ m/s, limited to the non-relativistic case. Then, according to equation (7), we get an estimate of the characteristic acceleration time of $\tau_{Mg} = \tau_{Fe} \approx 17$ days. The values of $\tau$ for iron and magnesium coincide because the density is inversely proportional to the mass of the dust particle [see eq. (7)], and, according to dependence (6), the acceleration of the spacecraft does not depend on the mass of the captured dust particle.

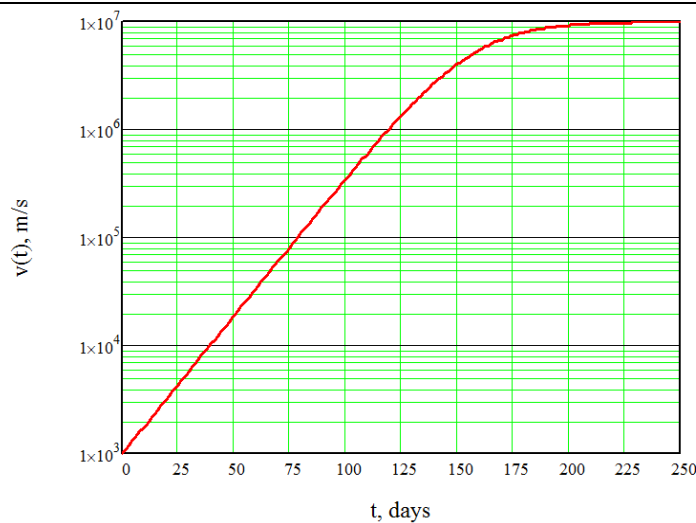

**Fig. 4.** The dependence of the spacecraft's speed v in m/s on the acceleration time in days.

This estimate shows that it is possible to obtain a reasonable acceleration time of the spacecraft, even in the non-relativistic case. Fig. 4 illustrates the dependence of the spacecraft's speed on the acceleration time for the initial parameters. As is seen from these graphs, the initial speed is increased by an order of magnitude over about one month. The maximum value of $v_{ex}$ is reached within six months.

## 6. The energy balance

This scheme is deemed feasible if the created plasma is accelerated to the required energies. We will estimate the ionization energies of neutral dust particles ionization process (Mg, Fe, Si atoms, specifically), which have the highest percentage in the interplanetary dust, by using the dependence of the unit power of Solar radiation, shown in Fig. 1. Moreover, these components are included in approximately equal parts [27—29]. To simplify calculations, we assign the highest ionization potential Si ($\varepsilon_i$ = 8.15 eV) to all dust particles. Instead of considering all the dust components, we introduce a particle with mass $\mu_\Sigma = \mu_{Mg} + \mu_{Si} + \mu_{Fe}/3 \approx 60 \times 10^{-24}$ g, where $\mu_{Mg}$, $\mu_{Si}$, $\mu_{Fe}$ - molecular masses of Mg, Si, and Fe, respectively. In the general case, the plasma flow can be accelerated to relativistic speed and the final kinetic energy of accelerated particles should be determined by the ratio

$$K = \mu_\Sigma c^2 \left( \frac{1}{\sqrt{1 - \left(\frac{u_{ex}}{c}\right)^2}} - 1 \right).$$

Although in the present estimates we consider only the non-relativistic case by limiting $u_{ex} \leq 10^7$ m/s.

For simplicity, we will only take into account elastic frontal collisions with the surface of the spacecraft, assuming that all the energy of dust particles is spent to decelerate the spacecraft. In this case, the energy balance can be written as

$$W_C S_* \tau_\varepsilon - S v \mu_\Sigma n \tau_\varepsilon \frac{v^2}{2} = \varepsilon_i N_0 + N_0 K/\zeta, \qquad (10)$$

where $S_*$ is the square of a solar panel, which are installed on the surface of the trap (here, for simplicity, it is assumed to be equal to the area of the entrance hole of the trap S, shown in Fig. 2), $\tau_\varepsilon$ is the accumulation time of the required solar energy. Also, it is worth noting that only a certain part of the solar radiation energy directly goes to the acceleration of particles, since there are losses on solar batteries and during ionization and acceleration processes. We use the energy utilization coefficient $\zeta$ to consider this effect parametrically. Taking into account

$$S v n \tau_{acc} = N_0,$$

from (10) we get

$$W_C S \tau_\varepsilon = N_0 \left( \varepsilon_i + K/\zeta + \frac{\tau_\varepsilon}{\tau_{acc}} \frac{\mu_\Sigma v^2}{2} \right). \qquad (11)$$

We can neglect the last term on the right side (11) for the case $\tau_\varepsilon/\tau_{acc} \ll 1$, since we consider the non-relativistic case of ship motion. Then from (11) we get the desired estimate

$$\tau_\varepsilon = \frac{N_0}{W_C S} (\varepsilon_i + K/\zeta). \qquad (12)$$

Moreover, taking into account $\zeta < 1$ and $\varepsilon_i \ll K$, we conclude that the influence of the ionization potential on the value of $\tau_\varepsilon$ is small, i.e. it does not depend on the type of ionized particle. In this case, the characteristic time of energy accumulation $\tau_\varepsilon$ is determined primarily by the speed of the spacecraft, the power of solar radiation, and the square of solar panels.

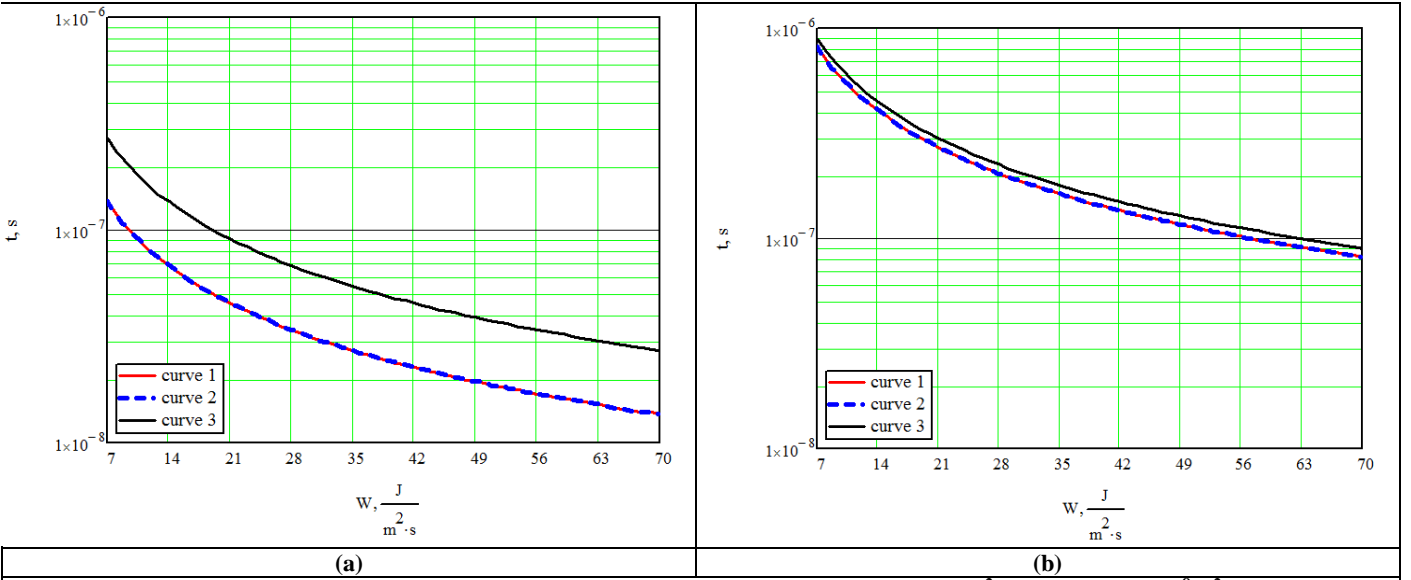

**Fig. 5.** The energy accumulation time $\tau_\varepsilon$ as a function of unit power of solar radiation $W_c$ [J/(m²s)] for $S$ = 3.14×10⁸ м² and final speed of the spacecraft $u_{ex} = 10^7$ м/с. Curve 1 corresponds to spacecraft's speed = $10^3$ m/s, curve 2 - $v = 10^5$ m/s and curve 3 - $v = 10^7$ m/s.

However, it should be borne in mind that these estimates of the time $\tau_\varepsilon$ should be considered an upper limit. This is how one needs to interpret these results shown in Fig. 5(a). They demonstrate the dependence of the accumulation time $\tau_\varepsilon$ on the unit power of solar radiation $W_C$ [J/(m²s)] for the distance between Venus and the asteroid belt for three different values of the spacecraft's speed v with the final speed of the dust $u_{ex} = 10^7$ m/s. In these estimations, we take the square of solar panels of $S* = 3.14 \times 10^8$ m². Curves 1, 2, and 3 correspond to the final speed of the spacecraft $v = 10^3, 10^5, 10^7$ m/s, respectively. As is seen, there is a small difference between curves 1 and 2 because the energy is mainly spent on an acceleration of particles to the final speed of $u_{ex} = 10^7$ m/s. The difference between Fig. 5(a) and Fig. 5(b) is due to the factor $\zeta$, which value in this study is assumed to be 0.1. This corresponds to the case when only 10% of the total absorbed radiation energy is converted to the acceleration.

The magnitude of the pulse directly depends on the amount of captured interplanetary medium, which follows from a comparison of the plots shown in Fig. 4 and 5. In such conditions, it is impossible to maintain a continuous operation mode of plasma engines, so pulse operation mode can be a good alternative. If the time of the pulse decreases by increasing the area of the trap, the dimensions of the spacecraft will grow. The increase of spacecraft's speed will reduce the time needed for accumulation particles and energy, so continuous operation mode could be possible for high speeds.

**Conclusion**

In the present model, we have studied the use of the interplanetary substance as fuel for plasma thrusters. We ignored the contribution of the solar wind to simplify estimates and limited our considerations only to stationary interplanetary dust in the region of $r < 2$ A.U. Consequently, the results of the study should be considered as a lower rough estimate, which demonstrates the possibility of technical implementation of the discussed scheme. We used astrophysical data to find the minimum and the maximum densities of the interplanetary dust $n_{Fe} = 2.32$ cm⁻³ and $n_{Mg} = 5.41$ cm⁻³, relatively. It is impossible to use the interplanetary medium as fuel for plasma thrusters for low speed of the spacecraft and such low densities, due to the low-pressure plasma source requires a density of the order $n_c \geq 10^{10}$ cm⁻³. Therefore, it was proposed to use the trap, shown in Fig. 2, to increase the density in the volume of the discharge chamber. In the example above (see Fig. 2), the trap increases the particle density irrespective of the spacecraft speed in k = $10^{10}$ times only due to geometric sizes. It has been demonstrated that the use of such a trap leads to an additional capture of interplanetary particles when there is an increase in the spacecraft speed. This allows reaching the required density of the ionized medium $n_0$ in the volume of the discharge chamber for a reasonable time (see Fig. 3).

In this study, solar radiation was considered as an energy source for the ionization of captured particles and their further acceleration. Using the power distribution of solar radiation (see Fig. 1), we have estimated the time needed to accumulate energy (see Fig. 5). As seen from the above dependencies, the incoming particle flow can only provide the pulsed operation mode for the plasma source, even using modern technics. In this case, the pulse frequency of such a device depends on the spacecraft speed and the density of the interplanetary medium.

Also, we took into account the energy losses connected to an implementation of a low-pressure RF discharge, the method of plasma acceleration, and the transformation of solar energy inside solar cells, using the coefficient $\zeta$ and we equated it to the value of 0.1. It seems that the value of $\zeta$ can be increased, and the estimates should be considered only as preliminary. They demonstrate only the principal feasibility of implementing the discussed scheme.

This work was supported by the Ministry of Science and Education of the Russian Federation under Grant 14.575.21.0169 (RFMEFI57517X0169).

**References**

[1] G. P. Sutton and O. Biblarz, *Rocket Propulsion Elements* (Wiley& Sons, New York, 2010).

[2] A. I. Morozov, *Introduction to plasma dynamics* (Fizmatlit, Moscow, 2008) [in Russian].

[3] S. D. Grishin, L. V. Leskov and N. P. Kozlov, *The Electric Rocket Propulsion* (Mashinostroenie, Moscow, 1975) [in Russian].

[4] R. G. Jahn and E. Y. Choueiri, *Electric Propulsion, in Encyclopedia of Physical Science and Technology* (Academic Press, New York, 2002).

[5] V. P. Kim, Tech. Phys. **60**, 362 (2015).

[6] A. N. Kozlov, J. Plasma Phys. **74**, 261 (2008).

[7] Y. Raitses, E. Merino and N. J. Fisch, J. Appl. Phys. **108**, 093307 (2010).

[8] A. R. Karimov and P. A. Murad, IEEE Trans. Plasma Sci. **46**, 882 (2018).

[9] O. A. Gorshkov, V. A. Muravlev and A. A. Shagaida, Hall-effect thrusters and plasma propulsion engines for spacecraft, (Mashinostroenie, Moscow, 2008) [in Russian].

[10] R. W. Bussard, Acta Astronaut. **6**, 1-14 (1960).

[11] R. W. Bussard and R. D. DeLauer, *Fundamentals of Nuclear Flight* (McGraw-Hill, New York, 1965).

[12] B. N. Cassenti, J. Br. Interplanet. Soc. **46**, 151 (1993).

[13] L. A. Singh and M. L. R. Walker, Prog. Aerosp. Sci. **75**, 15 (2015).

[14] L. Johnson, in Invited presentation at the Space Power and Energy Sciences International Forum (SPESIF) (Huntsville, Alabama, 2009).

[15] H. Blatter and T. Greber, Am. J. Phys. **85**, 915 (2017).

[16] A. A. Kosmodemyansky, *Theoretical Mechanics Course, Part II* (Prosveshchenie, Moscow, 1965) [in Russian].

[17] S. B. Pikel'Ner, Annu. Rev. Astron. Astrophys. **6**, 165 (1968).

[18] J. P. Valle, Fund. Cosmic Phys. **19**, 1-89 (1997).

[19] C. Leinert, E. Richter, B. Pitz and B. Planck, Astron. Astrophys. **103**, 177 (1981).

[20] I. M. Podgorniy and R. Z. Sagdeev, Usp. Phys. Nauk **98**, 410 (1969).

[21] J. A. Simpson, J. J. Connell, C. Lopate, R. B. McKibben, M. Zhang, J. D. Anglin, P. Ferrando, C. Rastoin, A. Raviart, B. Heber, R. Muiller-Meliin, H. Kunow, H. Sierks, G. Wibberenz, V. Bothmer, R. G. Marsden, T. R. Sanderson, K. J. Trattner, K. P. Wenzel and C. Paizis, Science 268, 1019 (1995).

[22] M. I. Pudovkin, ISSEP **12**, 87, (1996) [in Russian].

[23] I. Mann, A. Czechowski, N. Meyer-Vernet, A. Zaslavsky and H. Lamy, Plasma Phys. Control. Fusion **52,** 124012 (2010).

[24] M. Kivelson and C. Russell, *Introduction to Space Physics* (Cambridge University Press, Cambridge, 1995).

[25] P. Brady, T. Ditmire, W. Horton, M. L. Mays and Y. Zakharov, Phys. Plasmas **16,** 043112 (2009).

[26] L. Beirman, Usp. Phys. Nauk **90**, 163 (1966).

[27] E. GrËun, *Encyclopedia of the Solar System* (Academic Press, Massachusetts, 2006).

[28] E. N. Mironova, Dust in the solar system. Preprint at http://preprints.lebedev.ru/wp-content/uploads/2015/10/1015.pdf (2015) [in Russian].

[29] H. A. Zook, *Accretion of Extraterrestrial Matter Throughout Earth's History* (Springer, Boston, 2001).

[30] Yu P. Raizer, *Gas Discharge Physics* (Springer, New York, 1991).

[31] F. F. Chen, Plasma Phys. Control. Fusion **33**, 339 (1991).